\documentclass[reprint, superscriptaddress, amsmath, amssymb, aps, prx]{revtex4-1}
\usepackage{graphicx}
\usepackage{color}
\usepackage{graphicx}	
\usepackage{dcolumn}   	
\usepackage{bm}		    
\usepackage{float}

\DeclareMathOperator\erf{erf}

\usepackage{array}
\newcolumntype{C}[1]{>{\centering\let\newline\\\arraybackslash\hspace{0pt}}m{#1}}

\begin{document}
\title{Observation of Non-equilibrium Motion and Equilibration in Polariton Rings}
\author{S. Mukherjee}
\affiliation{Department of Physics and Astronomy, University of Pittsburgh, Pittsburgh, PA 15260, USA}
\author{D. M. Myers}
\affiliation{Department of Physics and Astronomy, University of Pittsburgh, Pittsburgh, PA 15260, USA}
\author{R. G. Lena}
\affiliation{Department of Physics and SUPA, University of Strathclyde, Glasgow G4 0NG, United Kingdom}
\author{B. Ozden}
\affiliation{Department of Physics and Astronomy, University of Pittsburgh, Pittsburgh, PA 15260, USA}
\affiliation{Department of Physics and Engineering, Penn State Abington, Abington, PA 19001, USA }
\author{J. Beaumariage}
\author{Z. Sun}
\affiliation{Department of Physics and Astronomy, University of Pittsburgh, Pittsburgh, PA 15260, USA}
\author{M. Steger}
\affiliation{National Renewable Energy Lab, Golden, CO 80401, USA}
\author{L. N. Pfeiffer}
\author{K. West}
\affiliation{Department of Electrical Engineering, Princeton University, Princeton, NJ 08544, USA}
\author{A. J. Daley}
\affiliation{Department of Physics and SUPA, University of Strathclyde, Glasgow G4 0NG, United Kingdom}
\author{D. W. Snoke}
\affiliation{Department of Physics and Astronomy, University of Pittsburgh, Pittsburgh, PA 15260, USA}
\begin{abstract}
We present a study of the macroscopic dynamics of a polariton condensate formed by non-resonant optical excitation in a quasi-one-dimensional ring shaped microcavity. The presence of a gradient in the cavity photon energy creates a macroscopic trap for the polaritons in which a single mode condensate is formed. With time- and energy-resolved imaging we show the role of interactions in the motion of the condensate as it undergoes equilibration in the ring. These experiments also give a direct measurement of the polariton-polariton interaction strength above the condensation threshold. Our observations are compared to the open-dissipative one-dimensional Gross-Pitaevskii equation which shows excellent qualitative agreement.   
\end{abstract}
\maketitle
\section{Introduction}
Exciton-polaritons in microcavities are now ubiquitous for studying fundamental properties of Bose-Einstein condensation \cite{SnokePT,LittlewoodUBEC}. In many ways, polariton condensate studies are complementary to cold atom condensate studies \cite{Jin1997,Stamper1998,Marago2000,Chevy2002,Moritz2003}, with continuous decay due to leakage through the confining mirrors of the microcavity allowing direct, time-resolved, non-destructive monitoring of the polariton condensate by external optics. Although this decay tends to also deter equilibration, in recent years it has been possible to make this term small enough that equilibrium condensation can be reached \cite{mitprl,Caputo2017}. Studies of polariton condensates near equilibrium, however, have so far been restricted to steady-state conditions \cite{Caputo2017,Gao2018,caputo2019josephson}. In this work, we present results for microcavity polaritons with ultra-long lifetimes ($\approx$ 200 ps) and transport distances up to hundreds of microns, which allow us to see macroscopic dynamics of a condensate all the way from a quench event to equilibrium; in particular, as the condensate nears equilibrium, we observe damped oscillations of a polariton condensate corresponding to pendulum motion modified by interaction effects in the condensate. These observations show that the kinetic and interaction energies play a significant role in the temporal dynamics, which we observe directly in real-time movies of the condensate. These studies also allow us to extract quantitative information about the polariton-polariton interaction constant, and lay the foundation for future studies of one-dimensional motion, including networks of paths with nodes and thermalization of strongly interacting gases.

In these experiments, two-dimensional (2D) planar microcavity structures, similar to those used in previous studies \cite{mitprl} were etched to make one-dimensional (1D) quantum wires in a ring geometry, as shown in Figure 1(b). This restricted the motion of the polariton condensate to 1D motion, while not decreasing their lifetime too much; as shown in previous work,\cite{Myers-apl} strain effects push the polaritons away from the free, etched surfaces, so that surface recombination does not play a significant role. As seen in Figure 1(c), the  potential-energy profile in the radial direction approximates a harmonic potential, giving a ladder of confined states in the radial direction. In the azimuthal direction, each of these confined states allows continuous 1D motion. The polaritons also experience an overall gradient of their potential energy due to a wedge in the optical cavity; as discussed in earlier work,\cite{steger-turn} this gradient leads to an effective ``gravity'' for the polaritons, demonstrated in macroscopic ballistic parabolic motion of the polaritons.  Because of this analogy to gravitational force, in this paper we will refer to the point of highest potential energy on the ring as the ``top'' of the ring, and the point of lowest potential energy as the ``bottom'' of the ring. The combination of the restriction to 1D circular motion and the overall potential-energy gradient gives rise to dynamics similar to rigid pendulum, up to modifications from interactions between condensate polaritons.

\begin{figure*}
\centering
\includegraphics[width =\linewidth]{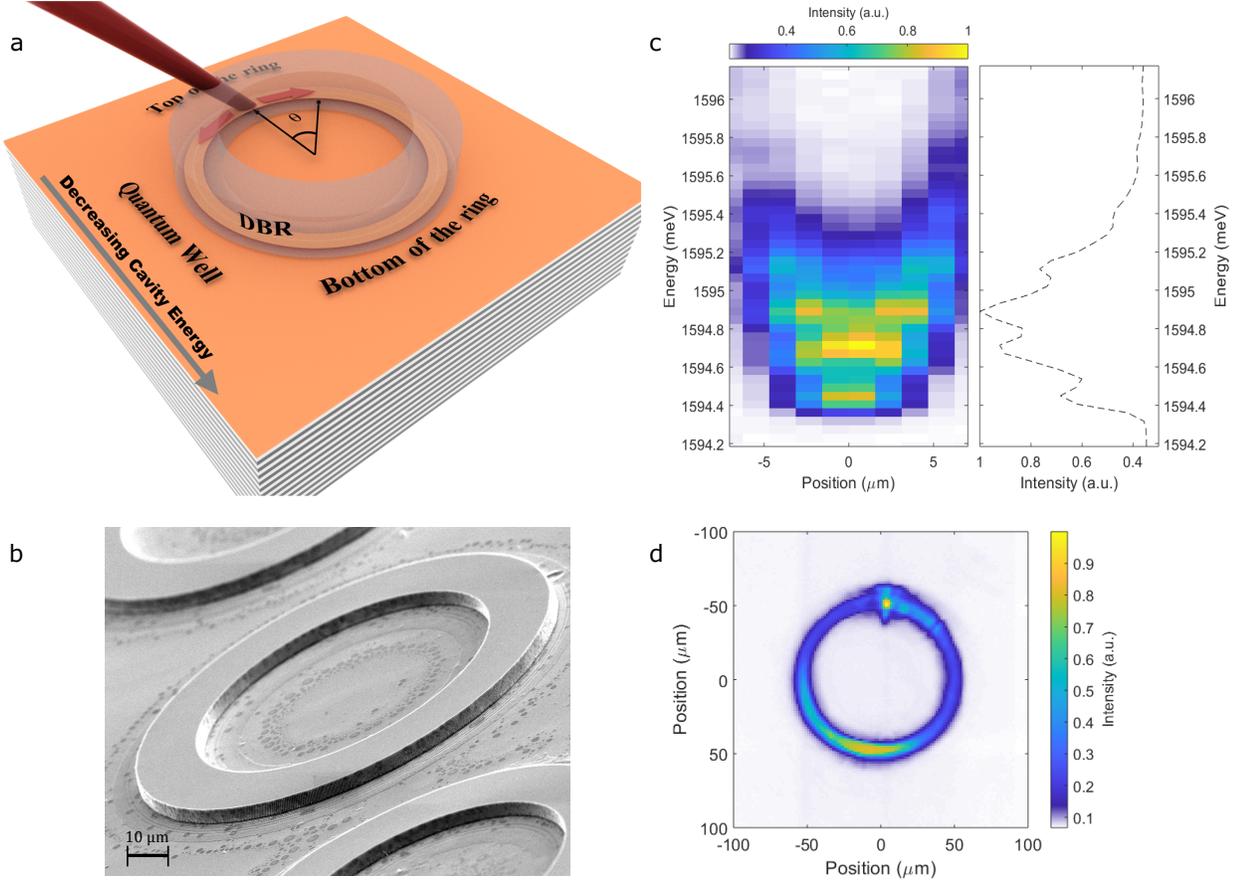}
\caption{\textbf{Ring channel microcavity.} a) A sketch of the ring structure showing the oblique incidence of the pump laser and introducing `top' and `bottom' nomenclature in the ring, and the definition of the angle ($\theta$) which is used to refer to different regions in the ring. b) Scanning electron microscope image of an etched ring used in the experiment. c) Radial mode separation in the ring channel with a width of 15 $\mu$m. d) Time-integrated and normalized photoluminescence from a typical ring structure for pulsed, non-resonant laser excitation at the top ($\theta =0$) of the ring. } 
\end{figure*}


\section{Experiment}
The polariton condensate was created by a short (few picosecond) laser pulse, localized to a spot a few microns in size, at the top of the ring ($\theta = 0$ as defined in Figure 1(a)). As discussed in numerous previous papers (e.g., Refs. \cite{Myers-apl,nelson-prx}) the non-resonant laser light creates a cloud of excitons at the laser excitation spot, which then convert down into polaritons; although there is a continuous transition between polariton and exciton states, it is useful to treat these as two distinct populations, with the exciton population at higher energy known as the ``exciton reservoir.''\cite{MyersArxiv2018}  Above a critical threshold of pump power, the polaritons at the creation spot undergo Bose-Einstein condensation in a highly nonequilibrium state, which then ballistically stream out of the exciton spot. As seen in Figure 2(a), the polaritons then rapidly fill the entire ring in a macroscopic condensate state far from equilibrium. (A movie of the motion of the polaritons in the ring is presented in the Supplementary information of this paper; Figure 2(a) gives single snapshots from this movie; Figure 1(d) shows time-integrated photoluminiscence from the ring.) 


\begin{figure*}
\centering
\includegraphics[width = \linewidth]{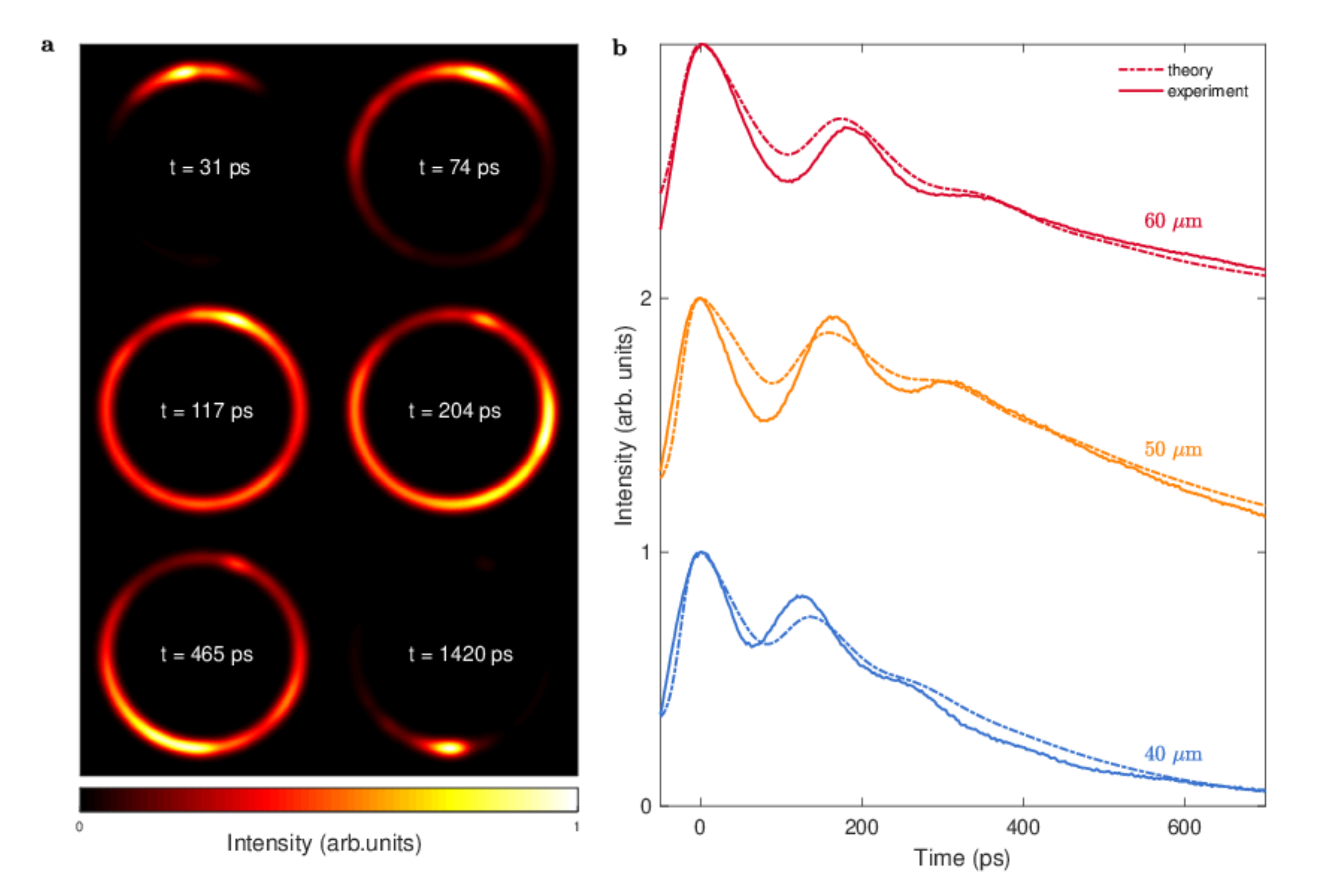}
\caption{\textbf{Polariton propagation in the rings and time period of oscillations.} a) Time snapshots of the PL from the ring with width 20 $\mu$m and radius 50 $\mu$m recreated by combining temporal intensity data from various locations on the ring with an angular resolution of 10$^{\circ}$. Each time frame has been normalized with respect to the maximum intensity from the ring in that frame. For a movie of the evolution, see the online Supplementary file. b) \textbf{Solid lines}: Intensity oscillations at the bottom of the rings as a function of time, for three different ring radii and with width 15 $\mu$m. \textbf{Broken lines}: Numerical simulation of the density evolution at the bottom of the ring due to the rigid pendulum potential.} 
\end{figure*}


As seen in Figure 2(a), and also in the movie, after the initial explosive, ballistic motion to fill the ring in the first 100 ps, the condensate then undergoes a much slower equilibration both spectrally and spatially into the bottom of the ring. Although the lifetime of the polaritons is around 200 ps, the time scale for this equilibration is much longer, because there is an interplay between the polariton population and the exciton reservoir with much longer lifetime. A previous study \cite{MyersArxiv2018} has shown that excitons created by the pump laser diffuse about 30~$\mu$m away from the creation spot, but here we see evidence of reservoir excitons all the way around the ring, 150~$\mu$m from the creation spot. This indicates that polaritons may convert back into excitons via polariton-polariton collisions. There are two effects of this background population of excitons. One is to extend the lifetime of the polaritons, as they can convert into excitons, with a lifetime of nanoseconds or greater, and then back into polaritons. The other effect of the exciton population is to give a smooth renormalization of the potential energy profile of the ring trap, due to the repulsion of polaritons from the excitons. This latter effect is seen in the shift to lower energy of the polariton emission spectrum at late times in Figure 2(b), as the background exciton population decays. 

\section{Pendulum Oscillations}
As seen in Figure 2(a), and also in Figure 2(b), after about 200 ps, when the polaritons are near equilibrium in the trap, pendulum-like oscillations arise. These oscillations are damped over a period of several hundreds of picoseconds and are seen in rings of different radii and widths. The period of these oscillations at low density is in good agreement with the natural frequency predicted for the classical low-amplitude oscillation frequency of a rigid pendulum, namely $\omega = \sqrt{g/R}$, where $R$ is the radius of the ring and $g$ is the effective gravity constant, given by $ mgh = \Delta U$, where $m$ is the effective mass of the polaritons, and $\Delta U$ is the potential energy difference for a distance $h$. A comparison of the time-resolved PL intensity for three different ring radii is shown in Figure 2(b). (For details of the optical pumping conditions, see Appendix.)  

Numerical simulations of the temporal and spatial dynamics for a polariton wave packet at late times are plotted with broken lines on top of the respective experimental data in Figure 2(b), for typical ring parameters given in the Appendix. 
The numerical model consists of a 1D Gross Pitaevskii equation (GPE) with generation $G(t)$ and loss $\Gamma(t)$ terms,
\begin{widetext}
\begin{equation}
    i\hbar\frac{\partial\psi(\theta,t)}{\partial t} = \left[-(1-i\alpha)\frac{\hbar^{2} \bigtriangledown^{2}}{2m} + V(\theta) +g_{1D}|\psi(\theta,t)|^2 +iG(t) -i\Gamma(t) \right]\psi(\theta,t),
\end{equation}
\end{widetext}
where $\psi(\theta,t)$ is the wavefunction for the condensate, $V$($\theta$) is the potential-energy profile of the ring and $g_{1\mathrm{D}}$ is the effective one dimensional polariton-polariton interaction strength. This model included some simplifications: we start with a wavefunction at the bottom of the ring, using the generation term $G(t)$ to account for the fast dynamics of polaritons moving from the top and considered a single radial mode decoupled to higher energy ones. For the numerical simulation, the above GPE was efficiently solved by using the split step Fourier technique \cite{antoine1,antoine2}.

Although this model has a few simplifications, we can draw two strong conclusions. One is that the oscillations are indeed driven by the interplay between the potential-energy profile and the kinetic energy term of Equation (1), which is sensitive to the long-range, spatial curvature of the condensate wave function. The second is that a kinetic-energy damping term, given by $i\alpha$ here in Equation (1), is  crucial to account for the damping of the oscillations as observed in the experiment. Such a term has been introduced by Malpeuch and coworkers in previous studies of polariton condensates \cite{barenghi2014}, and has been justified in the cold-atom literature by ZNG theory \cite{griffin2009} as due to the interplay between the condensate and thermal background particles. In the present experiments, this term may arise due to the interplay between the polariton condensate and the exciton reservoir. There may also be an interaction of the condensate with free carriers, as suggested by prior work \cite{hartwell} on the thermalization processes of polaritons.

Future work, to be published elsewhere, will include the coupling between different radial confined states and explicit tracking of the exciton reservoir population, to model the early-time behavior of the condensate. At present, we can already use these data to extract quantitative information about the polariton-polariton interaction strength $g_{1D}$. As discussed above, the period of the pendulum oscillations depends on the mass of the particles, unlike the case of a real gravitational pendulum, because the potential-energy gradient that gives the restoring force does not depend on the mass in this case, while the kinetic energy does. Therefore we can immediately conclude that the background excitons, which have mass very different from the polaritons, do not contribute to the oscillations, and only contribute a slowly varying, overall potential energy shift of the polaritons.

\begin{figure*}
\centering
\includegraphics[width = \linewidth]{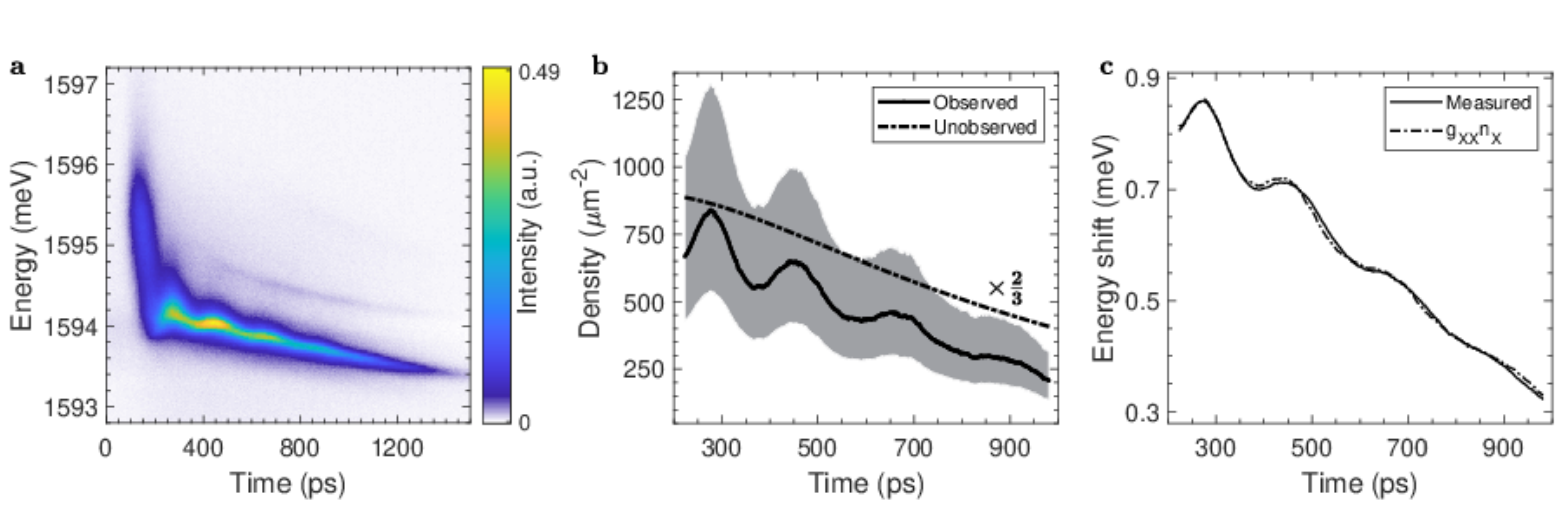}
\caption{\textbf{Measurement of the exciton-exciton interaction strength.} a) Energy-resolved photoluminescence intensity from the bottom of the ring ($\theta=180^\circ$) as a function of time following a short (2~ps) laser pulse at the top of the ring ($\theta = 0$), integrated over a wide angle of emission, corresponding to in plane $k$-vectors of the polaritons $-15^\circ < k_{\rm \|} < + 15^\circ$. The approximate level of the pump power used to excite the ring is 19 times the threshold power for condensation at the source, as seen in the transport to the bottom in this ring. b) \textbf{Solid lines:} Oscillations in the density of observed polaritons multiplied with the appropriate exciton fraction yielding an effective density of the excitons. The shaded area is the uncertainty in the density originating from the error bounds in the measurement. (Details in the Appendix). \textbf{Broken line:} Estimate of the exciton density invoked to obtain good fits to the energy shift using the mean field approach. c) \textbf{Solid line:} Oscillations in the energy shift of the ground state measured from respective streak images. \textbf{Broken line:} Energy shift deduced from the local density of the polaritons and excitons shown in Figure 3(b), for the best fit case.}

\end{figure*}

We can then compare the oscillations seen in two types of data to extract the polariton-polariton interaction parameter. The first is the oscillation of the polariton population at the bottom of the ring, which is observed in the emission intensity of the polaritons. Using the known calibration of photon emission rate to population density, presented in the Appendix, we can measure the density $|\psi(\pi,t)|^2$ to within $40\%$ of its absolute value. In addition, we have direct observation of the energy shift due to the renormalization of the polariton ground state energy, seen in the oscillation of the spectral energy position of the polariton photoluminescence, as shown in Figure 3(a). Assuming that the energy renormalization is linear with the polariton density, as expected for a mean-field energy shift, the comparison of these two oscillation magnitudes therefore allows a direct measurement of the constant $g_{1D}$, as shown in Figure 3(c). This number can then be converted to a value for the pure exciton-exciton interaction. The details of this analysis are presented in the Appendix, and give a value of the pure exciton-exciton interaction in a single quantum well of the order of 10 $\mu$eV-$\mu$m$^{2}$. This is in agreement with theoretical calculations \cite{Tassone}, and also in the same range as previous experimental measurements \cite{Estrecho2018,mitprl} when the polaritons are at high density, that is, at or above the condensation threshold, as is the case here. It is much lower than prior values reported for polaritons at low density \cite{natPhys2017, Ferrier2011, walker-prl}, indicating the possibility that the polariton-polariton interaction strength may depend on their density. 

\section{Conclusion}
 The results presented here, which are the only experiment with polaritons to reproduce the damped population oscillations seen in cold atoms, have been made possible by three experimental steps forward: first, the very high Q-factor of the microcavities used, which gives very long lifetime to the polaritons, allowing them to evolve in time all the way from a highly non-equilibrium, quenched initial state to spatial and spectral equilibrium at late times; second, the low disorder in these structures which allows long-range motion of the polariton condensate; and the development of etching processes which allows us to constrain the polariton motion to one dimension. The pendulum-like oscillations are not strongly affected by the interaction between the condensate polaritons in the regime observed here and at late times damped out to produce a fully thermalized, single-mode polariton condensate.

These dynamics are reproduced by a numerical solution of a Gross-Pitaevskii equation which includes generation and continuous decay, as well as energy renormalization due to the polariton-polariton interaction and kinetic energy loss due to interplay with the background population of thermal excitons. Future work will examine the coupling of different radial modes, which can be important at early times after the creation of the polaritons when the density of the polaritons is not too high; at high density, the condensate quickly thermalizes into just the lowest radial state.

By measuring the absolute density of the polaritons and their energy shift, we have extracted the polariton-polariton interaction strength, which is in agreement with other measurements at high particle density; all of the measurements in which there is a condensate give similar values for this parameter, while the outliers that give much higher interaction strength are from experiments with much lower polariton density.

Now that we can create 1D wires for polaritons with macroscopic transport, a natural extension is to create networks of one-dimensional wires for transport of coherent polaritons over macroscopic distances, i.e., polariton circuits. Because polaritons can be deflected by excitons and/or other polaritons, such circuits could allow all-optical switching networks of coherent light.
 
 \section{Acknowledgement}
 The work at Pittsburgh was funded by the Army Research Office (W911NF-15-1-0466). The work of sample fabrication at Princeton was funded by the Gordon and Betty Moore Foundation (GBMF-4420) and by the National Science Foundation MRSEC program through the Princeton Center for Complex Materials (DMR-0819860). Work at Strathclyde was supported by the EPSRC Programme Grant DesOEQ (EP/P009565/1). S. M. also acknowledges the support of the Pittsburgh Quantum Institute.
 \bibliography{refnat1}
 
 \appendix
 \setcounter{table}{0}
\renewcommand{\thetable}{T\arabic{table}}
 
 \setcounter{figure}{0}
\makeatletter 
\renewcommand{\thefigure}{S\@arabic\c@figure}
\makeatother

 \section{Sample details and experimental procedures}
 The top mirror of the microcavity structure used in this experiment consists of 32 alternating layers of Al$_{0.2}$Ga$_{0.8}$As and AlAs, while the bottom mirror consists of 40 alternating layers of the same materials. The entire structure was grown on a GaAs substrate by molecular beam epitaxy (MBE) with three sets of four 7 nm thick GaAs quantum wells with AlAs barriers embedded at the three antinodes of the cavity photon mode. The large Q-factor ($\sim 10^{6}$) of this microcavity results in long-lived polaritons with the lifetime of the order of 200 ps. Due to the spatial gradient in the rate of deposition of the layers during the MBE process, there is a thickness variation of the cavity. This leads to a strong variation of the cavity photon energy ($\approx$ 6$-$9 meV/mm) across the microcavity wafer. 

The gradient of the cavity also leads to a variation of the detuning, defined as the mismatch $\delta=E_{C}-E_{X}$ between the cavity photon energy $E_{C}$ and the quantum well exciton energy $E_{X}$. Since the detuning regulates the relative photonic and excitonic properties of the polaritons, we can have control of the lower polariton (LP) mass and the strength of the polariton-polariton interactions simply by examining rings at different places on the wafer. Typically the detuning is nearly constant over a single ring, while varying more significantly from ring to ring. 

Ultraviolet photolithography followed by high-density Cl$_{2}$/BCl$_{3}$ plasma etch was used to pattern the top mirror of the microcavity as rings of width (outer radius $-$ inner radius) 15 $\mu$m and 20 $\mu$m, and radius (average of outer and inner radius) of 40 $\mu$m, 50 $\mu$m and 60 $\mu$m. Figure 1(c) shows a typical example of the quantized transverse states, separated by about 260 $\mu$eV in this case, of a wire with 15 $\mu$m width. As in any quasi-1D structure, the separation of the states increases as the wire is made narrower. As seen in Figure 1(c), the energy states are nearly evenly spaced, as the transverse energy profile is nearly but not quite harmonic. A mode-locked Ti:sapphire pump laser with a pulse repetition of 76 MHz, a pulse width of $\approx 2$ ps, and a spot size of $\approx 20$ $\mu$m at an incident angle of $\approx 45^\circ$ was used for non-resonant excitation ($E_\mathrm{pump} \approx 1750$ meV) of the rings at their highest energy region. A 0.5 m spectrometer was used to spectrally resolve the photoluminescence (PL) with a 1200 grooves/mm reflective grating, and time-resolved measurements were done using a streak camera mounted on an exit port of the spectrometer. PL from the rings was collected using a microscope objective with a numerical aperture (NA) of 0.40. All measurements were performed by cooling the microcavity to low temperature (below 10 K) in a continuous-flow cold-finger cryostat. 

To count all the photons entering our collection optics and reaching the CCD of the streak camera, we used a half wave plate and a polarizer before the entrance slit of the spectrometer. For all the measurements the orientation of the transmission axis of the polarizer was kept fixed, which removed the polarization sensitivity of all the optics inside the spectrometer as the light that entered the spectrometer was always at a fixed linear polarization. Two images were taken for each individual measurement, one with the half-wave plate fast axis (placed just before the final polarizer) at 0$^{\circ}$ and one with it at 45$^{\circ}$. By adding these two images together, the total contribution of both polarizations (both parallel and orthogonal to the final polarizer) were taken into account. The setup was calibrated to convert the intensity counts on the CCD to the photon counts emitted from the ring. The photon counts were subsequently converted into an absolute polariton number per state after accounting for the lifetime of the LP state. Details about this calibration can be found in the Appendix for this paper. This method has been shown previously to agree accurately with density measured independently from chemical potential of the polariton gas at the Bose-Einstein condensation threshold \cite{mitprl}. 

\section{Experimental parameters for observation of pendulum oscillations}

The pump power used to observe pendulum oscillations in PL intensity as in Figure 2(b) of main text was approximately 6 times the threshold power for condensation at the source, as seen in the transport to the bottom in these rings, which is not sufficient to drive the polaritons at the bottom into a single transverse mode condensate. Therefore the dominant contribution to the PL intensity comes from higher order transverse and non-zero in-plane momentum k$_\|$ states. Measurement of the ring parameters are given in Table T1.

\begin{table*}[htbp]
\centering
\begin{tabular}{c c c c c}
\hline
R ($\mu$m) & $\Delta$U (meV) & m & $2\pi\sqrt{\frac{R}{g}}$ (ps) & $T_{\mathrm{meas}}$ (ps) \\
\hline
60 & 0.88 & $6.0\times10^{-5}$m$_{0}$ & 331 & 343 $\pm$ 5 \\
50 & 0.83 & $6.4\times10^{-5}$m$_{0}$ & 302 & 306 $\pm$ 5 \\
40 & 0.66 & $6.7\times10^{-5}$m$_{0}$ & 270 & 272 $\pm$ 5\\
\hline
\end{tabular}
\caption{Comparison of time periods for different rings studied in Figure 2(b). $\Delta$U is found by taking a slice of the PL across a diameter of the ring under a broad, low power pump spot. The uncertainty in this measurement is approximately 10$\%$. Mass of the LP is found by fitting a parabola to the LP dispersion for $k_{\|}$ close to zero. m$_{0}$ is the free electron mass in vacuum (5.1$\times10^{5}$eV/c$^{2}$). $T_{\mathrm{meas}}$ is the time period of the oscillations measured between the first and the third peak from the experimental data.}
\label{tab:TimePeriod}
\end{table*}

\section{Example of streak images used for measuring interaction strength}

Energy resolved streak images of the PL from the bottom of the ring are shown in the first column of Figure S1. The detuning at the bottom of the ring in Figure S1(d-f) which is a re-plot of Figure 3 of main paper is $-4.7$ meV, while for Figures S1(a) and S1(g) it is $-4.1$ meV, which leads to a slight variation in the LP mass between these two rings. The pump power used to capture these streak images is much higher (by approximately a factor of two or more) than the previous condition discussed above. In this high density regime, we measured the time period of the oscillations to be longer than what is predicted from the energy gradient in the rings. This is due to the reshaping of the potential at the bottom of the ring,  which makes the effective gravity weaker in the rings. The effect of the nonlinearity arising due to the interactions between the polaritons on the dynamics can be understood as a renormalization of the mass of the lower polaritons. In all of these figures, we see clear oscillations in the intensity as well as in energy with time. At early times we see a significant population in non-zero in-plane momentum states ($k_{\|}<2$ $\mu \mathrm{m}^{-1}$), which corresponds to the arrival of the spatially inhomogeneous precondensate from the top of the ring. As the population at the bottom builds up, the occupation in the finite $k_{||}$ states is depleted while the emission from the $k_{\|}=0$ state increases. This is also reflected in the spectral narrowing of the emission with time. The oscillations in the intensity of the emission are related to the dynamics of the condensate in a rigid pendulum potential. As the energy of the emission also tracks the rise and fall of the intensity, which is expected from a mean field approximation, we apply this theory to directly measure the exciton-exciton interaction strength $g_{xx}$ as outlined below. A natural advantage of measuring the repulsive interaction strength in this experiment over other steady-state experiments using the non-resonant pump is that there are constraints on the measured density and energy which come from the dynamics of the system, resulting in tighter bounds on the measured value of the interaction strength. 

\begin{figure*}
\centering
\includegraphics[width = \linewidth]{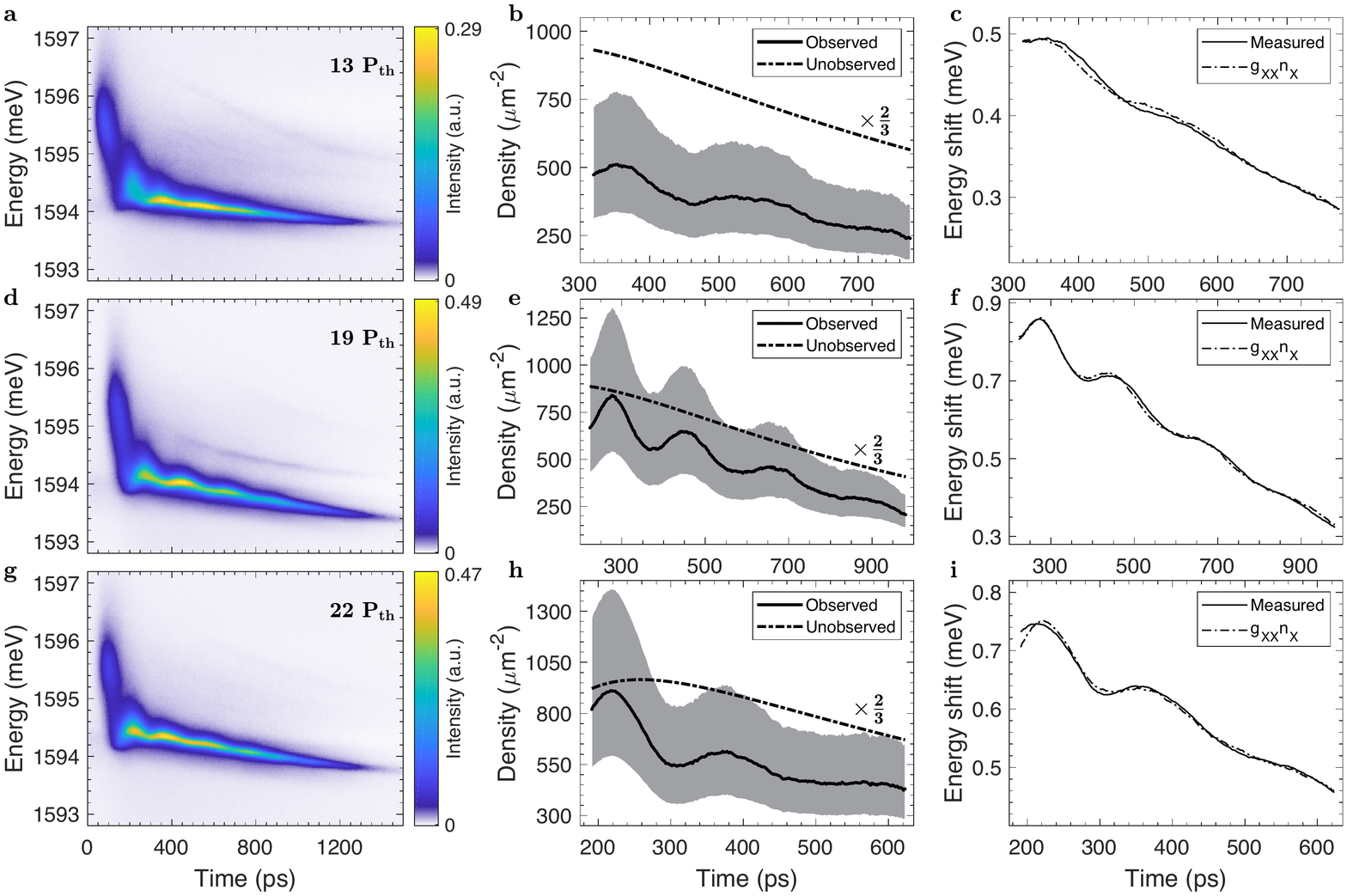}
\caption{a, d, g) Energy-resolved PL intensity from the bottom of the ring as a function of time following a short (2~ps) laser pulse at the top of the ring. The value reported in the top right corner in each of these plots indicate approximate level of the pump power used to excite the rings. b, e, h) \textbf{Solid lines:} Oscillations in the density of observed polaritons multiplied with the appropriate exciton fraction yielding an effective density of the excitons. The shaded area is the uncertainty in the density originating from the error bounds in the measurement. \textbf{Broken lines:} Estimate of the exciton density invoked to obtain good fits to the energy shift using the mean field approach. c, f, i) \textbf{Solid lines:} Oscillations in the energy shift of the ground state measured from respective streak images. \textbf{Broken lines:} Best fit to the observed energy shift using the mean field formula by using the total density of the excitons.}

\end{figure*}

Mean field theory predicts that the local blue shift in the energy in a condensate having repulsive particle-particle interactions is proportional to the local density of the particles. It therefore seems quite straightforward to directly fit the observed evolution of the ground state at the bottom of the ring to the density evolution of the observed polaritons at the same point in the ring. However, we find poor fits for all the data that we have analyzed if we use this method alone. An example is shown in Figure S2(b), for a $50$ $\mu$m ring radius with 15 $\mu$m etch width. The oscillations in both the density and the energy are correlated in phase and period, but the amplitude of the oscillations is relatively less in the blue shift than in the density. This indicates that something else is giving rise to an additional blue shift besides the polariton-polariton interaction.

\begin{figure*}
\centering
\includegraphics[width = \linewidth]{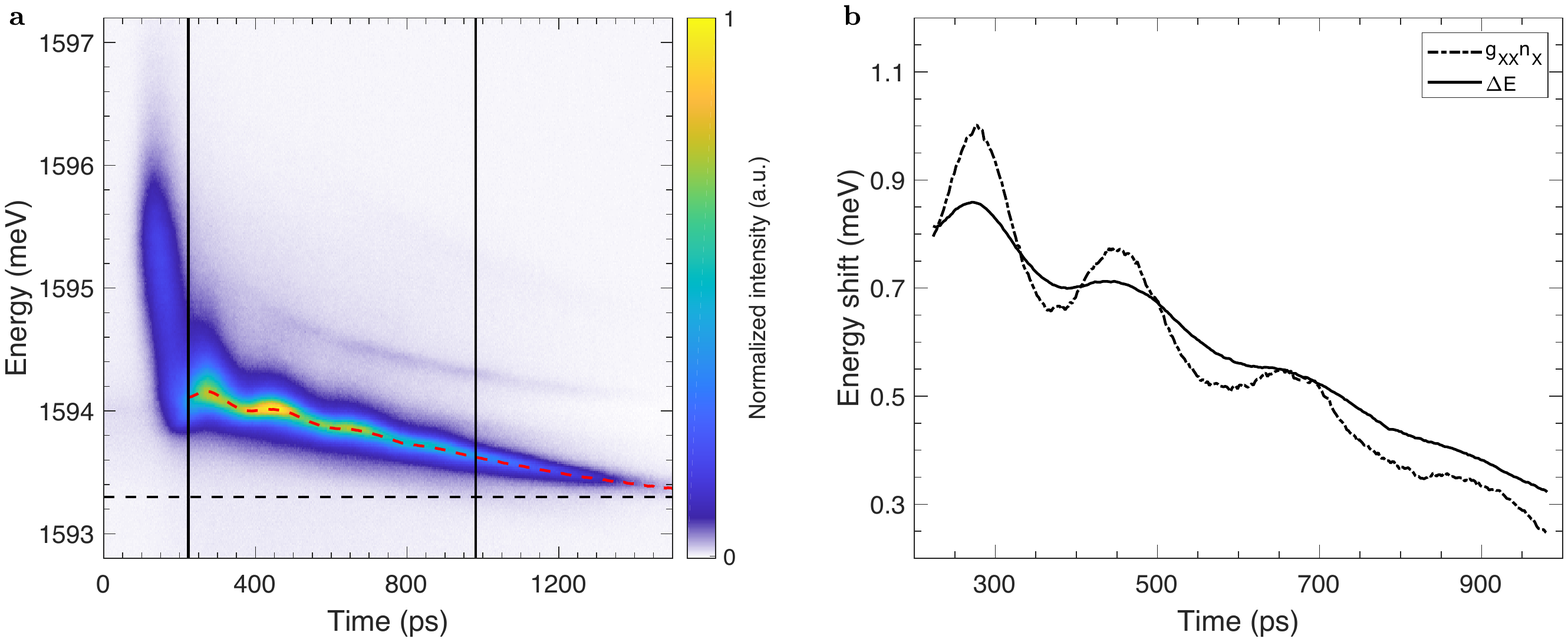}
\caption{Example of a fit with no exciton reservoir. a) Re-plot of Figure S1(d) with red dashed lines showing the extracted energy oscillations using a Lorentzian linewidth. Black dashed lines show the reference ground state energy measured at 1593.3 meV at the bottom of this ring. Energy shift at early times is measured with respect to this value. Black solid lines show the temporal window used for fitting the density and energy oscillations. b) Black solid curve is the measured energy shift of the ground state from Figure S2(a). Broken lines is the best fit of the energy shift obtained from the observed polariton density and assuming no contribution from the exciton reservoir. The fit returns a value for $g_{XX}=39.9$ $\mu$eV$\mu$m$^{2}$ which lies within the 95$\%$ confidence interval of (39.5, 40.3) $\mu$eV$\mu$m$^{2}$ .}
\end{figure*}

As we have seen above, the oscillation frequency is dependent on the lower polariton mass near the bottom of the ring consistent with the rigid pendulum model. Therefore the cause of the oscillations in both the density and the energy data can be attributed entirely to the local changes in the lower polariton density. A reasonable conclusion is therefore that the remaining, non-oscillating energy shift arises from the polaritons interacting with a background exciton population, which must also have moved around the ring from the laser excitation spot. The presence of the exciton population will not affect the time period of the oscillations because their mass is much greater than that of the polaritons. Motion of excitons away from the laser spot for distances of 20-30 $\mu$m has been directly verified by recent experiments \cite{MyersArxiv2018}. Here, at least some of the excitons must have traveled 150 $\mu$m around the ring from the pump spot, which is quite surprising, because excitons in GaAs quantum wells were previously observed to migrate only a few microns away from the laser excitation region. This indicates that the placement of the quantum wells in the optical cavity to give the strongly coupled polariton effect strongly affects the effective diffusion of the excitons in the quantum wells.

The second column of Figure S1 shows the effective exciton density due to polaritons alone, found by multiplying the measured polariton density by the respective exciton fraction. As discussed above, this density is not sufficient to explain the blue shift of the ground state energy of the polaritons by a linear proportionality. Therefore, we guessed a population evolution of the (unseen) excitons at the same location as the polaritons.
The simplest choice for the time dependence of this exciton population is to have a rise time and a fall time as given by 
\begin{equation}\label{eq:1}
    n_{r} = \tilde{n}_{r}\Big(1-e^{-(t-t_{1})/\tau_{1}}\Big)e^{-(t-t_{2})/\tau_{2}}.
\end{equation}
The parameters of this function were varied to give the best fits to the measured energy shifts, as shown in the third column of Figure S1, where the fit also gives the exciton-exciton interaction strength $g_{xx}$. The quality of these fits assessed from the R-squared estimator is above 99$\%$ for all the three fits. Knowing the absolute density of the polaritons from photon counting as discussed above, and the number of quantum wells (12) over which the excitons are distributed, we obtain the following values for the pure exciton-exciton interaction $g_{xx}$:  8.4 $\mu$eV-$\mu$m$^{2}$ with an uncertainty interval of (4.9, 14.2) for Figure S1(c), 13.5 $\mu$eV-$\mu$m$^{2}$ with an uncertainty interval of (8.4, 23.5)  for Figure S1(f), and 10.2 $\mu$eV-$\mu$m$^{2}$ with an uncertainty interval of (6.1, 17.8) for Figure S1(i). The uncertainty interval is determined by adding the exciton reservoir density to the upper and lower boundaries of the observed polariton density uncertainty (shown with shaded region in the second column of Figure S1) and fitting the total density to the observed energy shift. We determined the sensitivity of $g_{xx}$ on the reservoir density by varying the density of the reservoir used in each of these fits by changing $\tilde{n}_{r}$ within a reasonable range while keeping the R-squared estimator of the fits above 99$\%$. We found a drop in the value of the estimator with an increasing spread of the fit residues, indicating worse fits. Thus we conclude that the dynamics severely constraints our estimate of the reservoir density as well. 

\section{Finding the interaction strength from a streak image}

The energy and time-resolved image of the PL taken by the streak camera shows how the energy distribution of the polariton states at a given location in the ring evolve with time. To make a direct measurement of the interaction strength we determine the polariton density as a function of time from the image pixel counts. This process involves two steps. The first step is to relate the counts recorded at the pixel to the number of photons leaking from the cavity. This requires a careful calibration of our optical setup at the central wavelength of PL emission. The second step involves converting the photon counts to the number of polaritons inside the microcavity using the lifetime of the polaritons. From the streak image, we find that the polaritons have a spread in their in-plane momentum. Therefore to correctly convert, we must consider the in-plane momentum dependence of the lifetime of the polaritons, which makes this conversion slightly non-trivial. In this regard, the first step is to find a mapping between the energy of the polariton and its in-plane momentum, which is usually provided by the dispersion relation of the polaritons in the microcavity. We consider the effect of the interactions between the polaritons, which renormalises their dispersion relation adding an energy correction proportional to the density at the mean field approximation level. Since the density of the polaritons changes with time, so does the energy-momentum dispersion of the polaritons. 

The interactions between the polaritons originate from the exciton part and so for weak interactions, we expect the energy of the excitons to be blue shifted by $g_{\mathrm{xx}}n_{\mathrm{x}}$, where $g_{\mathrm{xx}}$ is the exciton-exciton interaction strength and $n_{\mathrm{x}}$ is the exciton density. In the absence of any reservoir of excitons ($n_{\mathrm{res}}$) this will be given by $f_{\mathrm{x}}n_{\mathrm{pol}}$, where $f_{\mathrm{x}}$ is the exciton fraction of the polaritons, defined as
$f_{\mathrm{x}}(k_{\|}) = \bigg(1+\delta(k_{\|})/\sqrt{\delta(k_{\|})^{2}+\Omega^{2}}\bigg)/2$ and $n_{\mathrm{pol}}$ is the density of the polaritons. Therefore the blue shift in the exciton energy ($\Delta E_{\mathrm{x}}$) can be expressed as
\begin{equation}
    \Delta E_{\mathrm{x}} \sim g_{\mathrm{xx}}\Big( f_{\mathrm{x}}n_{\mathrm{pol}} + n_{\mathrm{res}}\Big).
    \label{eq:sim}
\end{equation}
$\Delta E_{\mathrm{x}}$ can be determined by diagonalizing the standard coupled Hamiltonian for the photon and exciton modes both with and without the interaction shift. Given the LP energy defined by
$E_{LP} = \bigg[ (E_{C}+E_{X}) -  \sqrt{(E_{C}-E_{X})^{2}+\Omega^{2}}\bigg]/2$, the overall shift due to the change in the exciton energy is
\begin{widetext}
\begin{equation}
E'_\mathrm{LP}(0) - E_\mathrm{LP}(0) = \frac{\Delta E_{\mathrm{x}}}{2} + \frac{1}{2}\bigg(\sqrt{\delta^2 + \Omega^2 } -\sqrt{(\delta - \Delta E_{\mathrm{x}})^2 + \Omega^2}\bigg),
\label{eq:interactionShift}
\end{equation}
\end{widetext}
where $E'_\mathrm{LP}(0)$ is the interaction-shifted LP energy, and $E_\mathrm{LP}(0)$ is the low density LP energy, both at $k_\| = 0$. The Rabi splitting energy ($\Omega$) and the detuning ($\delta$) are determined from the sample characterization procedure. The detuning ($\delta$) used in the above equation is the strain-adjusted detuning in the ring channel. $E'_\mathrm{LP}(0)$ is directly measured at each time step from the streak image by fitting the $k_{\|}$ = 0 emission with a Lorentzian lineshape. $E_\mathrm{LP}(0)$ is found from the measurement of the LP energy under a broad and low power pump spot at the region of our interest. The above equation is numerically inverted to find $\Delta E_{\mathrm{x}}$ at each time step. It is straightforward to see that the new energy dispersion for the polaritons will have smaller radius of curvature near zone center and thus will have lighter mass. 

Polariton lifetime ($\tau_{\mathrm{pol}}$) is usually given by the cavity photon lifetime ($\tau_{\mathrm{cav}}$) from the relation
\begin{equation}
    \tau_{\mathrm{pol}} \approx  \frac{\tau_{\mathrm{cav}}}{f_{\mathrm{cav}}},
\end{equation}
where $f_{\mathrm{cav}}$ is the cavity fraction of the polaritons with $k_{\|}$ dependence given by
\begin{equation}
    f_\mathrm{cav}(k_\parallel) =\frac{1}{2}\bigg(1-\frac{\Delta E(k_\parallel)}{\sqrt{\Delta E(k_\parallel)^2 + \Omega^2}}\bigg).
\end{equation}
$\Delta E(k_\parallel)$ is defined as the energy difference between the cavity photon ($E_{\mathrm{cav}}$) and the exciton ($E_{\mathrm{exc}}$). For the range of $k_{\|}$ considered in our measurements, $E_{\mathrm{exc}}$ is essentially constant for varying $k_{\|}$ whereas $E_{\mathrm{cav}}\propto\sqrt{k_{\|}^{2}+k_{\perp}^{2}}$. The dependence of $E_{\mathrm{cav}}$ on $k_{\|}$ leads to the dependence of $f_{\mathrm{cav}}$ on $k_{\|}$. With $\Delta E_{\mathrm{x}}$ known as a function of time, we map the LP energy for each pixel of the image to corresponding $|k_{\|}|$ value using the renormalized dispersion relation. This will help to determine the number of polaritons ($N_\mathrm{pol}$) in the time interval (t, t + $\Delta t$) and energy interval (E, E + $\Delta \mathrm{E}$),
\begin{equation}
  N_\mathrm{pol} \approx \Delta N_\mathrm{phot} \frac{\tau_\mathrm{cav}}{f_\mathrm{cav}\Delta t},
\end{equation}
where $\Delta N_\mathrm{phot}$ is the number of photons counted by the camera pixel with energy between E and E + $\Delta \mathrm{E}$ in the time interval (t, t + $\Delta t$). Putting everything together the observed polariton density ($n_{\mathrm{pol}} = N_\mathrm{pol}/A$) is

\begin{equation}
  n_{\mathrm{pol}}(t) = \sum_{i}  \frac{\Delta N_\mathrm{phot}(E_{i},t)\tau_\mathrm{cav}}{A f_{\mathrm{rep}} \Delta T f_\mathrm{cav}(E_{i},t)\Delta t}, 
\end{equation}
where index $i$ runs over the energy bins, $f_{\mathrm{rep}}$ is the repetition rate of the laser and $\Delta T$ is the integration time of the image. As mentioned in the main text, the fits to the energy oscillations are poor when using only the observed polariton density. Therefore, we consider blueshift of the ground state ($k_{\|}=0$) per particle due to polaritons and reservoir interactions which leads to
\begin{widetext}
   \begin{equation}
\begin{split}
 \Delta E_{LP}(t) & = g_{\mathrm{pol-pol}}n_{\mathrm{pol}}(t) + g_{\mathrm{pol-ex}}n_{\mathrm{res}}(t) 
\\
& = \tilde{g}_{\mathrm{xx}}f^{\mathrm{0}}_{\mathrm{x}}\bigg(f^{\mathrm{0}}_{\mathrm{x}}n_{\mathrm{pol}}(t) + n_{\mathrm{res}}(t)\bigg)
\\
&\simeq \tilde{g}_{\mathrm{xx}}f^{\mathrm{0}}_{\mathrm{x}}\bigg(\sum_{i}\frac{f_{\mathrm{x}}(E_{i},t)\Delta N_\mathrm{phot}(E_{i},t)\tau_\mathrm{cav}}{A f_{\mathrm{rep}} \Delta T f_\mathrm{cav}(E_{i},t)\Delta t}+n_{\mathrm{res}}(t)\bigg)
\\
&= g_{\mathrm{fit}}\bigg(\sum_{i}\frac{f_{\mathrm{x}}(E_{i},t)\Delta N_\mathrm{phot}(E_{i},t)\tau_\mathrm{cav}}{A f_{\mathrm{rep}} \Delta T f_\mathrm{cav}(E_{i},t)\Delta t}+n_{\mathrm{res}}(t)\bigg),
\end{split}
\label{eq:sim2}
\end{equation} 
\end{widetext}
where $f^{\mathrm{0}}_{\mathrm{x}}$ is the low density, zero in-plane momentum exciton fraction given by $(1 + \delta/\sqrt{\delta^{2}+\Omega^{2}})/2$ and $g_{\mathrm{fit}}$ = $\tilde{g}_{\mathrm{xx}}f^{\mathrm{0}}_{\mathrm{x}}$. The term in parenthesis is the total effective density of excitons which is used to fit the energy shift in Figure 3 of main text and Figure S1 above. We considered the $k_{\|}$ dependence of the exciton fraction while calculating the effective density of the excitons from the observed polariton density. It is also noted that the above relation could be derived from Equation \ref{eq:sim} and Equation \ref{eq:interactionShift} within the mean field approximation. It is quite straightforward to show $\Delta E_{LP}(t)=f^{\mathrm{0}}_{\mathrm{x}}\Delta E_{\mathrm{x}}(t) + O(\Delta E_{\mathrm{x}}^{2})$ from Equation \ref{eq:interactionShift}. And using Equation \ref{eq:sim}, we find $\Delta E_{LP}(t)=f^{\mathrm{0}}_{\mathrm{x}}\tilde{g}_{\mathrm{xx}}\Big( f_{\mathrm{x}}n_{\mathrm{pol}}(t) + n_{\mathrm{res}}(t)\Big)$ after neglecting higher order terms in density. The value of $\tilde{g}_{\mathrm{xx}}$ obtained from Equation \ref{eq:sim2} is related to the total density observed for all the quantum wells in the microcavity. In order to make direct comparison with the theoretical prediction of $g_{\mathrm{xx}}$ in a single quantum well, we multiply the fit value of $g_{\mathrm{fit}}$ by the number of quantum wells $N_{QW}$ and divide by the appropriate exciton fraction. 

Additionally, one may consider a contribution to the blueshift from the saturation of the exciton oscillator strength at high exciton density in the quantum wells which result in the decrease of Rabi splitting energy \cite{Brichkin2011}. Since both the contributions are proportional to the exciton density, we cannot distinguish them using the measurement of $g_{\mathrm{fit}}$ using Equation \ref{eq:sim2} and $g_{\mathrm{fit}}$ will now be given by
\begin{equation}
    g_{\mathrm{fit}} = g_{\mathrm{xx}}f^{\mathrm{0}}_{\mathrm{x}} +  g_{\mathrm{sat}}\sqrt{f^{\mathrm{0}}_{\mathrm{x}}(1-f^{\mathrm{0}}_{\mathrm{x}})}.
\end{equation}

\begin{figure}
\centering
\includegraphics[width = \linewidth]{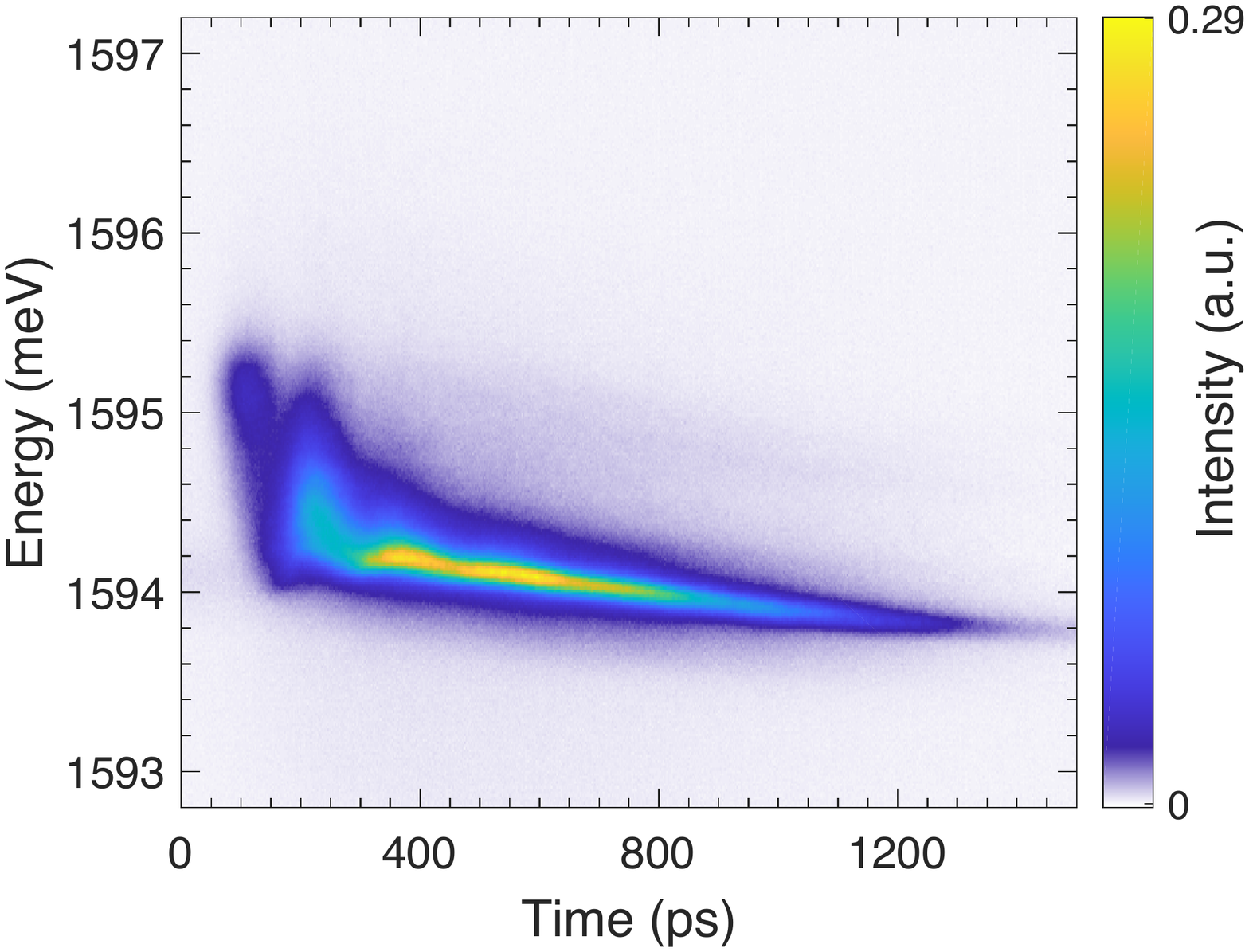}
\caption{Photoluminescence intensity for the same conditions as that of Figure S1(a), but with the range of in-plane $k_{\|}$ collected in the experiment cut down to $-2^\circ < k_{\rm \|} < + 2^\circ$, using an aperture in the collection optics, corresponding to the emission from just the ground state of the ring. }
\end{figure}

\begin{figure*}
\centering
\includegraphics[width = 0.8\linewidth]{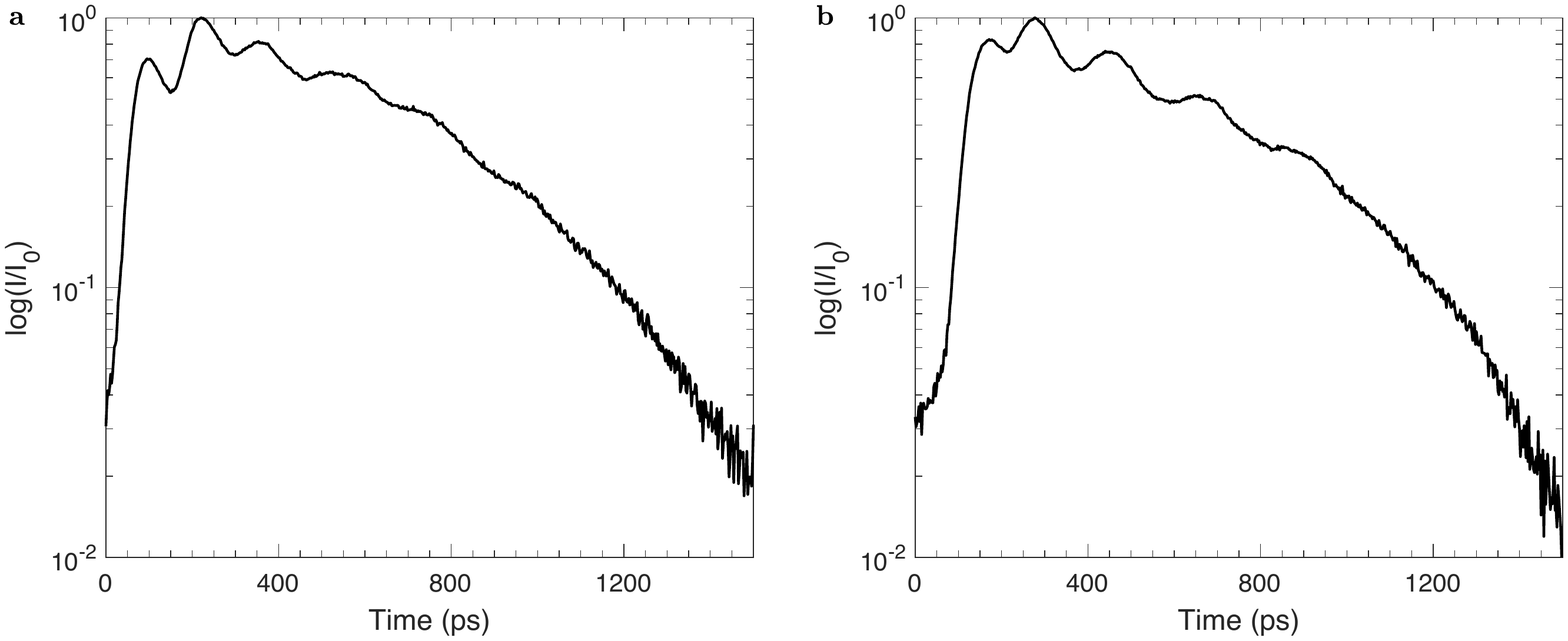}
\caption{Normalized PL intensity plotted on a log-linear scale. (a) corresponds to the streak image shown in Figure S1(g) while (b) corresponds to the streak image shown in Figure S1(d). We do not observe any single exponential decay of the signal for regions with signal-to-noise ratio $>$ 0.1.}
\end{figure*}

\section{Determining error bounds on $g_{\mathrm{fit}}$}
The uncertainty in the fit parameter $g_{\mathrm{fit}}$ arises from the measurement of the polariton density. Explicitly this depends on the intensity counts at each pixel in the CCD camera, the conversion factor ($\mathrm{\eta}$) to convert intensity counts to photon counts, the cavity lifetime, the absolute cavity fraction and the area of the region from where the PL has been measured. Uncertainty in the intensity recorded by the pixels of the CCD are related to the random noise which follows a normal distribution with full width half maximum of 22 counts. $\mathrm{\eta}$ is the photon collection efficiency of our setup, which was determined by sending a known power of light at the same wavelength as the polariton PL through the setup from the location of the sample, and then imaging it in the same way that the data is collected during an actual experiment. For our setup and typical image parameters, this was 151 $\pm$ 11 photons/count. The cavity lifetime (135 $\pm$ 10 ps) used in this study was previously estimated for a similar microcavity structure\cite{steger-turn}. We could not establish a good estimate for the polariton lifetime from the streak images because it did not show a single exponential decay, even for very late times (see Figure S4), which is an indication for the presence of an exciton reservoir.

Together the spectrometer slit and the time slit of the streak camera selected a rectangular region from the image of the PL which corresponded to a collection area of 53.94 $\pm$ 3.72 $\mu \mathrm{m}^{2}$ on the sample. The uncertainty in determining the absolute cavity fraction is related to the uncertainty in the sample characterization as discussed in the Appendix of Ref. \cite{MyersArxiv2018}. These uncertainties are used to find an upper and lower bound on the observed polariton density as shown by the shaded area in Figure S1. Another source of uncertainty arises from the choice of the unobserved exciton reservoir density. We extensively varied the density of the reservoir to determine the best fits to the energy shift. The density of the exciton reservoir shown in Figure S1 corresponds to these best fit cases. It must be noted that any error in part of finding the energy shift due to the uncertainty in finding the ground state energy of the polaritons is compensated by the assumption of invoking a reservoir density and the value of $E_{LP}(0)$ only determines an offset of the energy, without affecting the oscillations. As a demonstration of this notion we compare the fit obtained in Figure S1(f) with the fit where the measured energy shift is offset by 1 meV. By choosing an appropriate reservoir density we obtained best fit with nearly identical value of the fit parameter (see SI Figure S5). The uncertainty in measuring $E'_{LP}(0)$ at a given time slice obtained from the 95$\%$ confidence interval of the Lorentzian fits are small ($<$ 50 $\mu$eV) and is therefore neglected.  
\begin{figure*}
\centering
\includegraphics[width = \linewidth]{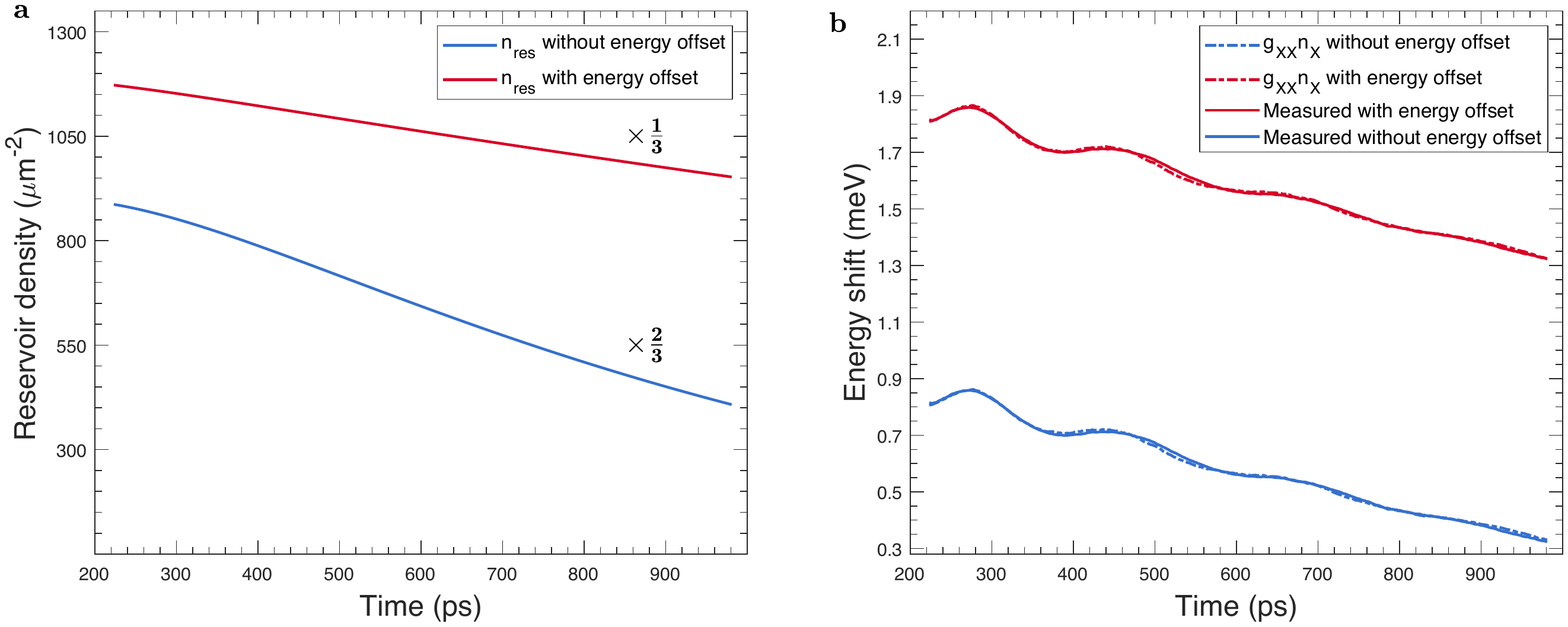}
\caption{Comparison of fits with and without an energy offset to $E_{LP}(0)$. (a) The red curve shows the new density of the reservoir used to obtain a good fit as shown in (b). The blue curve is the re-plot of the reservoir density from Figure S1(e) in main text. (b) Fit to the measured energy shift with (red solid lines) and without (blue solid lines) an energy offset of 1 meV using the observed polariton density (as plotted in Figure S1(e)) and the reservoir density in (a). The value of $g_{XX}$ obtained from the new fit is 14.5 $\mu$eV$\mu$m$^{2}$ which is within the error bounds $(8.4, 23.5)\mu$eV$\mu$m$^{2}$ as reported in the main text. This shows that any constant energy shift due to the presence of reservoir does not affect our measurement of $g_{XX}$ due to the constraints from the dynamical feature in the data.}
\end{figure*}

\section{Theoretical model}
To model the oscillations in the PL from the condensate at the bottom of the ring as shown in Figure 2(b) in main text, we employ the open-dissipative Gross-Pitaevskii (ODGPE) model \cite{wouters2007} with a phenomenological kinetic damping term $\alpha$, a spatially homogeneous generation term with temporal rise and fall behavior G(t) and an effective loss term $\Gamma$(t) to account for the overall decay of the total population:
\begin{widetext}
\begin{equation}
    i\hbar\frac{\partial\psi(\theta,t)}{\partial t} = \left[-(1-i\alpha)\frac{\hbar^{2} \bigtriangledown^{2}}{2m} + \frac{V_0}{2}(1-\cos(\theta)) +g_{1D}|\psi(\theta,t)|^2 +iG(t) -i\Gamma(t) \right]\psi(\theta,t),
\end{equation}
\end{widetext}
where $\theta \in$ [$-\pi,\pi$] and $\theta$ = 0 is the lowest energy point in the ring. We choose $G(t)$ of the form $G_{0}(1 - exp(-t/\tau_{1}))exp(-t\tau_{2})$ to approximately match the rise time of the PL intensity and $\Gamma(t)$ of the form $\Gamma_{0}(1 - exp(-t/\tau_{3}))$ to match the particle loss rate at later times. We note that the inclusion of the kinetic damping term not only damps the motion of the condensate about the bottom of the trap, but also results in additional particle loss. We correct for this by appropriately renormalizing the wavefunction in our calculation to obtain only the overall loss implied by the terms in $G(t)$ and $\Gamma(t)$. It is useful to transform the above equation into its dimensionless form by introducing $x_{d}=x/a_{0}$, $t_{d}= \omega t/2\pi$ and $\psi_{d}=\sqrt{a_{0}/N}\psi$, where $a_{0}=\sqrt{\hbar/m\omega}$ and $\omega=\sqrt{V_{0}/2mR^{2}}$ are the natural length scale and the natural angular frequency of the harmonic oscillator potential associated with the bottom of the ring potential respectively. We also choose to work on a linear geometry by unwrapping the co-ordinates on the ring using the relation $x=R\theta$. The dimensionless equation then takes the form
\begin{widetext}
\begin{equation}
    i\frac{\partial\psi_{d}}{\partial t_{d}} = \left[-(1-i\alpha)\pi\frac{\partial^{2}}{\partial x_{d}^{2}} + \frac{\pi V_0}{\hbar\omega}(1-\cos(a_{0}x_{d}/R)) +\frac{2\pi g_{1D}N}{\hbar\omega a_{0}}|\psi_{d}|^2 +iG_{d}(t_{d}) -i\Gamma_{d}(t_{d}) \right]\psi_{d},
\end{equation}
\end{widetext}
where $x_{d} \in[-\pi R/a_{0},\pi R/a_{0}]$, $G_{d}$ and $\Gamma_{d}$ are the appropriate dimensionless gain and loss functions. We choose a dimensionless gaussian wavefunction with a variance of $\sigma$ centered at $\theta$ = 0 with unity norm defined on the domain of $x_{d}$ as the initial condition 
\begin{equation}
   \psi_{d}(x_{d},0)=\frac{1}{\sqrt{\sigma\sqrt{\pi}\erf(\pi R/\sigma a_{0})}}\exp^{-x_{d}^{2}/2\sigma^{2}}.
\end{equation}
This initial condition is propagated forward in time according to the dimensionless ODGPE defined above with $g_{1\mathrm{D}}N$ = 0.25 $\mu$eV-mm and the parameters for the generation and loss functions with the damping coefficient as listed in Table S1.
\begin{table*}[htbp]
\centering
\begin{tabular}{c c c c c c c c}
\hline
R ($\mu$m)   &   $\omega\tau_{1}/2\pi$   &   $\omega\tau_{2}/2\pi$   &   $\omega\tau_{3}/2\pi$   &   2$\pi G_{0}/\hbar\omega$   &   2$\pi\Gamma_{0}/\hbar\omega$   &   $\alpha$   &   $\sigma$\\
\hline
60 & 0.20 & 0.6 & 0.5 & 1.0 & 1.0 & 0.8 & 5\\
50 & 0.20 & 0.6 & 1.5 & 1.0 & 1.0 & 1.1 & 6 \\
40 & 0.15 & 0.6 & 0.5 & 1.0 & 0.9 & 1.1 & 5\\
\hline
\end{tabular}
\caption{Parameter values used in the simulation for Figure 2(b) in the main text.}
\label{tab:Parameters}
\end{table*}

The results from this simulation (in Figure 2(b)) match reasonably well with the oscillation amplitudes with the typical ring parameters given in Table T1. We also compare these results with the linear case when the time evolution is only under the ring potential term and find that the time periods are in good agreement with each other as shown in Figure S6. This underlines the dominant role played by the ring trap potential in determining the oscillations period we observed here. 
\begin{figure*}
    \centering
    \includegraphics[width = 0.9\linewidth]{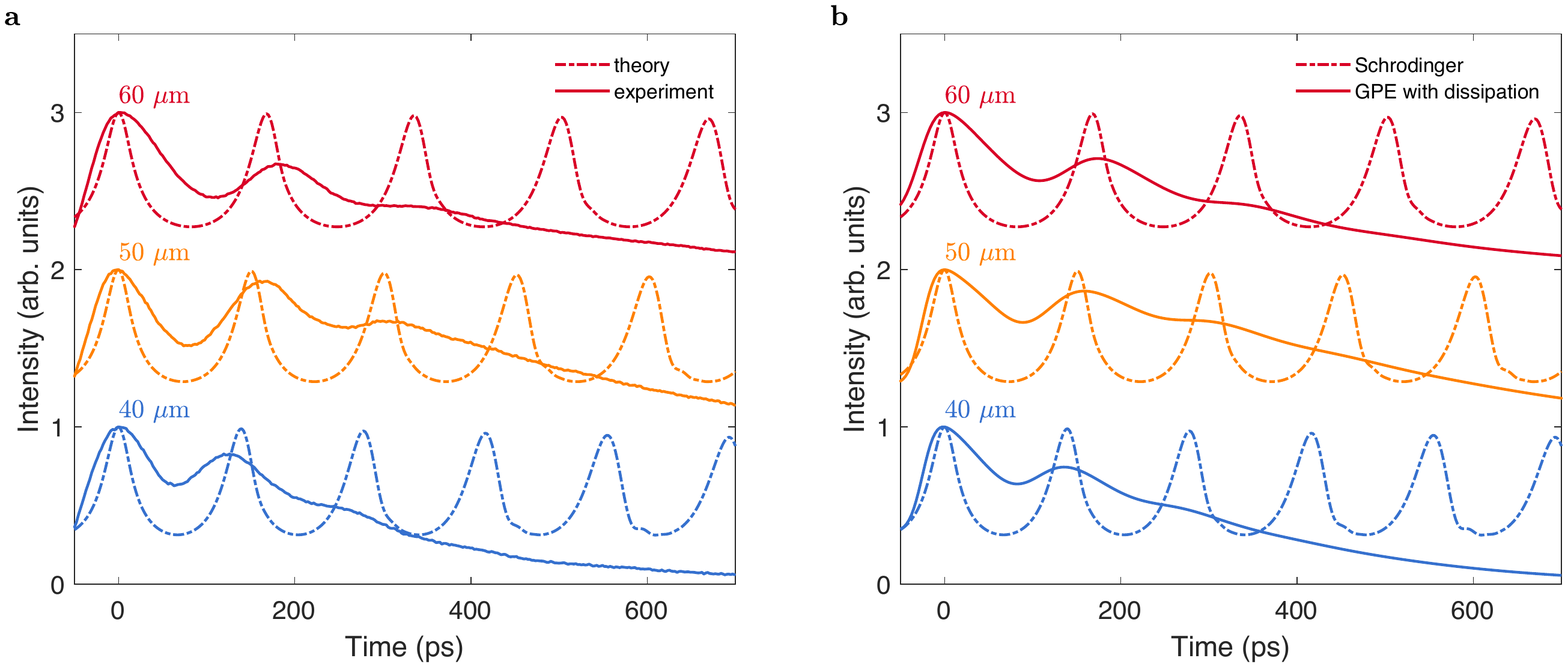}
    \caption{Time evolution of the polariton condensate density at the bottom of the ring. (a) Broken lines show the evolution under the Schrodinger operator and the solid line compares the observed PL intensity. (b) We make a direct comparison between the two theoretical models and show that the simpler model can capture the oscillation time period as long as the potential trap is not reshaped at higher polariton density. We also notice that, while when neglecting the interaction term the density peaks are narrower than the experimental ones, when turning the repulsive interaction on they become broader and their width increases with increasing interaction strengths. }
    \label{fig:SI4_2}
\end{figure*}
\end{document}